\documentclass[aps,prc]{revtex4}

\usepackage{epsfig}
\usepackage{dcolumn} 
\usepackage{amsfonts}
\usepackage{color}

\newcommand{\be}{\begin{equation}}
\newcommand{\ee}{\end{equation}}
\newcommand{\ba}{\begin{array}}
\newcommand{\ea}{\end{array}}
\newcommand{\bea}{\begin{eqnarray}}
\newcommand{\eea}{\end{eqnarray}}

\def\half{{1\over2}}
\def\Hb{\mathbb{H}}

\begin{document}


\title{A computationally tractable version of the collective
model}


\author{D.J.~Rowe}
\affiliation{Department of Physics, University of Toronto\\
Toronto, Ontario M5S 1A7, Canada}

\begin{abstract} 
A computationally tractable version of the Bohr-Mottelson collective model is
presented which makes it possible to diagonalize realistic collective models and
obtain convergent results in relatively small appropriately chosen subspaces of the
collective model Hilbert space.
Special features of the
 proposed model is that it makes use of the beta wave functions given analytically by
the softened-beta version of the Wilets-Jean model, proposed by Elliott {\em et
al.}, and a simple algorithm for computing SO(5) $\supset$ SO(3) spherical harmonics.
The latter has much in common with the  methods of Chacon, Moshinsky, and Sharp but
is conceptually and computationally  simpler.
Results are presented for collective models ranging from the sherical vibrator to
the Wilets-Jean and  axially symmetric rotor-vibrator models.
\end{abstract}
\maketitle 

\section{Introduction} 

This paper presents a computationally tractable
version of the standard Bohr-Mottelson (BM) collective model
\cite{BM,BMbook,Rowebook}. 
The need for such a version of the model arises because
the expansion of rotational wave functions in a spherical vibrational basis is so
slowly convergent that the diagonalization of a general coupled-rotor-vibrator
collective model Hamiltonian in this basis is virtually impossible. 
The proposed method constructs a basis for the collective model in which beta wave
functions, centred about a non-zero equilibrium value, are given analytically
following the methods of Elliott {\it et al.}\
\cite{EEP86} and ref.~\cite{RB98}, 
and new methods are developed for computing
the complementary rotational and gamma wave functions.
The motivation for this development is to be seen in the context of a
sequence of steps from a phenomenological model description of nuclear
collective structure to a microscopic interpretation of what it means.
I follow the strategy of giving a phenomenological
collective model a microscopic foundation by first formulating it in
algebraic terms and subsequently constructing representations of its
algebraic expression on a microscopic shell model Hilbert space.

Early advances in the shell model theory of collectivity came with the
identification of  seniority and symplectic symmetry with pairing
\cite{Flowers,Kerman,Talmi}  and su(3) 
 with rotations \cite{Elliott}.  An early
description of the use of symmetry in nuclear structure was provided by 
Parikh \cite{Parikh}.

Algebraic methods have subsequently had an enormous influence on collective model
theory (cf.\ \cite{Rev} for a review). 
Two influential developments have been: the Frankfurt
version of the collective  model (\cite{Greiner,Hess}) 
and the Interacting Boson Model (IBM) \cite{IBM}. 
The Frankfurt
methods were developed to give solutions to the collective model in the intermediate
region between the analytically solvable vibrational and rotational limits.
They made major use of the algebraic structures associated with the
five-dimensional harmonic oscillator.
The IBM achieved a reduction of the Frankfurt
program to a finite-dimensional space by compactifying the algebraic
structure of the collective model to the U(6) symmetry group of the
six-dimensional harmonic oscillator and restricting  consideration to single
(finite-dimensional) U(6) irreps. The IBM has three exactly solvable
limits, corresponding to similar limits of the BM model. 

Major developments towards the goal of formulating a microscopic
(shell model) theory of collective states have also followed the algebraic
approach (a historical survey was provided in the review article of ref.\
\cite{EarlyRev}). 
The latter developments were based on the symplectic
model \cite{RosRow}. 

The symplectic model and its several variations have not been applied widely
to fit detailed nuclear data because that was not their purpose (cf.\ refs.\
\cite{BR2000,CRKB} for some recent applications).  Their purpose was to obtain a
fundamental explanation of nuclear collective dynamics and the way it emerges from
interacting nucleons. Thus, even if one were to succeed in formulating a completely
satisfactory theory of nuclear collective dynamics, it would hardly be a simple
theory.  Nor would it serve the purpose of every-day analysis of nuclear data.
For this purpose, one continues to need simple phenomenological models, albeit
preferably ones with a microscopic foundation.
With this concern in mind, we revisit the BM collective model with the
benefit of insights acquired from the Frankfurt model, the IBM, and
the symplectic model.

\section{The standard solvable limits of the collective model}

The classic BM collective model shares a Hilbert space with the
five-dimensional harmonic oscillator. Indeed, in its harmonic vibrational
limit, its spectrum and eigenstates are precisely those of the
five-dimensional harmonic oscillator.
Thus, it has rich algebraic, geometrical, and analytical
structures all which are exploited in this paper.  It has a spectrum
generating algebra given by the semi-direct sum Lie algebra [HW(5)]u(5) and a
corresponding dynamical group [HW(5)]U(5).  The Lie algebra [hw(5)]u(5) is
spanned by five $(L=2)$ pairs of $d$-boson (phonon) operators
$\{ d^\nu, d^\dagger_\nu ; \nu = 0,\pm 1,\pm 2\}$
and the infinitesimal generators $\{ d^\dagger_\mu d^\nu\}$ of U(5), where
the $d$-bosons satisfy the commutation relations 
$[d^\mu,d^\dagger_\nu ] = \delta_{\mu\nu}$.

The collective model has three well-known algebraically solvable limits:
the harmonic vibrator model, the Wilets-Jean  (gamma-soft) model
\cite{WJ}, and the axially-symmetric rigid-rotor model.
These submodels are associated with dynamical subgroup
chains corresponding to different paths through the set of groups
\be \begin{array}{ccccccc}
{\rm [HW(5)]U(5)} & \supset & {\rm [R^5]SO(5)} & \supset &
{\rm [R^5]SO(3)}\\
\cup && \cup && \cup\\
{\rm U(5)} &\supset&{\rm SO(5)} &\supset&{\rm SO(3)} &&
\end{array} 
\ee
starting with [HW(5)]U(5) and ending with SO(3),
where R$^5$ is the group with Lie algebra spanned by the quadrupole
moments
\be \Big\{\hat Q_\nu = {1\over \sqrt{2}} \, (d^\dagger_\nu + d_\nu) \,, 
\nu = 0, \pm 1, \pm 2\Big\}
\ee
with $d_\nu = (-1)^\nu d^{-\nu}$.
Thus, 
\be
{\rm [HW(5)]U(5)}  \supset {\rm U(5)}\supset  {\rm SO(5)} \supset {\rm SO(3)} 
\label{eq:U5chain}\ee
is a dynamical subgroup chain for the harmonic vibrator model, 
\be
{\rm [HW(5)]U(5)}  \supset {\rm [R^5]SO(5)} \supset {\rm SO(5)}
\supset {\rm SO(3)}  \label{eq:WJchain}
\ee
is a dynamical subgroup chain for the  Wilets-Jean (beta-rigid,
gamma-soft) model, and
\be
{\rm [HW(5)]U(5)}  \supset {\rm [R^5]SO(5)} \supset {\rm [R^5]SO(3)}
\supset {\rm SO(3)}  \label{eq:RRchain}
\ee
is a dynamical chain for the rigid-rotor (beta- and gamma-rigid) model. 
The above subgroup chains are discussed in more detail,
for example, in ref.\ \cite{Rev}.

The solution of more general collective model Hamiltonians has been tackled, for
example, by Hess {\em et al.}\ \cite{Hess} in the  U(5) $\supset$ SO(5) $\supset$
SO(3) basis for the collective model Hilbert space of Chacon, Moshinsky and Sharp
\cite{CMS}. This approach has been very influential and has applications to the IBM
in which the same U(5) and SO(5) $\supset$ SO(3) groups also appear.
Its limitation is that it is impractical for
the description of collective model states of large deformation which converge
extremely slowly in a spherical U(5) basis.

\section{An alternative basis for the collective model}

Moving between the above solvable limits of the collective model is complicated by
the `rigidity' of two of the limits. Both the rigid-rotor model and the beta-rigid
gamma-soft (Wilets-Jean) model have delta function components to their wave
functions and, as a consequence, they are not realizable in the Hilbert space of the
vibrator limit except as limits of sequences of normalizable wave functions. This
limitation expresses the fact that rigidly-defined intrinsic quadrupole moments are
unphysical and incompatible with quantum mechanics (as well as relativity theory).
I therefore  consider an alternative basis for the diagonalization of collective
model Hamiltonians and show that it can lead to convergent solutions for a wide
range of collective model Hamiltonians.

 An important feature of the collective model is that its coordinates
separate into orthogonal subsets in much the same way as those of a single
particle in three-dimensional space separate into radial and spherical
coordinates. 
Associated with  this observation is the fact that the Hilbert space
of the collective model is a direct sum 
\be \Hb^{{\rm CM}} = \bigoplus_v \Hb_v= \bigoplus_v\, \Hb^{{\rm
SU(1,1)}}_v\otimes\Hb^{{\rm SO(5)}}_v\,,\label{eq:BMHilb}\ee 
of Hilbert spaces labeled by a seniority quantum number
$v$, which each carry an irrep of a direct product group SU(1,1)$\times$SO(5), where
SU(1,1) is the group of scale transformations of the beta (radial)
coordinate, defined by
$\beta^2 = Q\cdot Q$, and SO(5) is the five-dimensional rotation group.
As a result of this separation of variables, SO(5)-invariant collective model
Hamiltonians can be diagonalized with just the Lie algebra of SU(1,1) as
spectrum generating algebra.

Thus, for example, the collective model has two other exactly
solvable submodels given by simply adding a $1/\beta^2$ term to either the
five-dimensional harmonic oscillator Hamiltonian \cite{EEP86} or to the
five-dimensional analog of the Coulomb Hamiltonian \cite{FV}.
Such models are known in three dimensions as the Davidson \cite{Davidson}
and Kratzer models \cite{Krat}.
Algebraic solutions of three-dimensional central-force problems, have been
studied widely in terms of the algebra of the direct product group
SU(1,1)$\times$SO(3) (reviewed in ref.\ \cite{CP}; an historic account
is given in Wybourne's book \cite{Wyb} and an overview  of the basic
methods  in ref.\ \cite{CW}).
The natural extension of the algebraic treatment to five-dimensional space
with dynamical group SU(1,1)$\times$SO(5) is straightforward
\cite{RB98,Rev}. The fact that solvable central force problems
remain solvable on addition of a
$1/r^2$ term to the Hamiltonian follows from the observation that the
commutation relations of the SU(1,1) spectrum generating algebras are
unchanged,  by the substitution
$\nabla^2 \to \nabla^2 +k /r^2$. This result holds in 
higher-dimensional spaces.

Following the standard methods for three-dimensional central-force problems , the
spectrum and wave functions for  the five-dimensional harmonic oscillator
\be \hat H = -{\hbar^2\over 2B}\nabla^2 + \half B\omega^2 \beta^2 ,\ee
are given by
\be E_{nv} = (2n+\lambda_v)\hbar\omega, \quad n = 0, 1, 2, \dots
\quad v = 0, 1, 2, \dots \label{eq:engylevels}\ee 
and
\be \Psi_{nv\sigma}(\beta, \omega) = \sqrt{2n!\over \Gamma (n+\lambda_v)
b^5}\ \Big( {\beta\over b}\Big)^{\lambda_v - {5\over 2}}
\exp \Big( - {\beta^2\over 2b^2}\Big) \,
L^{(\lambda_v-1)}_n \Big(  {\beta^2\over b^2}\Big)\,
\mathcal{Y}_{v\sigma}(\omega) , \label{eq:basis}\ee
where $\mathcal{Y}_{v\sigma}$ is an SO(5) spherical harmonic for an SO(5)
irrep of seniority $v$, $L_n^{(\lambda_v-1)}$ is an associated Laguerre
polynomal, and $\lambda_v = v +5/2$, $b=\sqrt{\hbar/B\omega}$.

The above (standard) results are obtained 
by use of the SU(1,1) Lie algebra spanned by the operators
\be \hat Z_1 = -\nabla^2 , \quad \hat Z_2 = \beta^2 , \quad \hat Z_3 =
-i  (q\cdot \nabla + \textstyle{5\over 2})\,.
\ee
Because the commutation relations of these operator are
unchanged if $\hat Z_1$ is replaced by
\be \hat Z_1' =  -\nabla^2 + {\beta_0^4\over\beta^2},
\ee
it  follows \cite{RB98} that the harmonic vibrational limit of the
collective model extends immediately to a model with beta-vibrational Hamiltonian
\be \hat H(\beta_0) = {\hbar^2\over 2B}\Big(- \nabla^2 +{\beta_0^4\over
\beta^2}\Big)+ \half B\omega^2\beta^2 , \label{eq:betaHam}\ee
The only change in the results of the harmonic vibrational limit for the
energies $\{E_{nv}\}$ and wave functions $\{ \Psi_{nv\sigma}\}$ is that 
 the relationship $\lambda_v = v + 5/2$ is generalized to
\be \lambda_v = 1+\sqrt{(v + \textstyle {3\over 2})^2 + \beta_0^4} \,.
\label{eq:lambda}
\ee
This is because    
the replacement $\hat Z_1 \to \hat Z_1'$ in the SU(1,1) Lie algebra
results  \cite{RB98} in a modification of the value of the SU(1,1) Casimir invariant
from the value
$\lambda_v (\lambda_v -2) = v (v+3) + {5\over 4}$ to
the value
\be \lambda_v (\lambda_v -2) = v (v+3) + \textstyle{5\over 4} +
\beta_0^4\,.\ee 
The five-dimensional Kratzer model \cite{FV} is handled in a similar way as shown in
the Appendix.

With explicit wave functions, it is a simple matter to compute
observable properties of the model states.
For example, the matrix elements of an electric quadrupole operator of the form
\be \hat Q_M = Z \beta\, \hat q_M \label{eq:Qop}
\ee
with $\hat q_M$ taking the values
\be q_M =\cos\gamma \mathcal{D}^2_{0,M}(\Omega) + 
\frac{1}{\sqrt{2}} \sin\gamma \left(
\mathcal{D}^2_{2,M}(\Omega)+\mathcal{D}^2_{-2,M}(\Omega)\right) \,,
\label{eq:qop}\ee
factor into products of a matrix element of $\hat\beta$ between associated
Laguerre polynomials and a matrix element of a $v\!=\!1$ SO(5) spherical harmonic.
Thus, the latter matrix elements are given by elementary SO(5) Clebsch-Gordan
coefficients as shown explicitly below.

The properties of a sequence of collective models with Hamiltonian $\hat H(\beta_0)$
and with $\beta_0$ ranging from zero to a large value are discussed in the following
section.
It is seen there that the above model provides a sequence of analytically
solvable solutions which progress  from the harmonic vibrational 
${\rm [HW(5)]U(5)}\supset{\rm U(5)}$ limit to the
asymptotic Wilets-Jean  ${\rm [HW(5)]U(5)}\supset{\rm [R^5]SO(5)}$ limit.

Subsequent sections demonstrate  that
the basis  of eqn.~(\ref{eq:basis}), which reduces an
${\rm SU(1,1)}\times{\rm SO(5)}\supset{\rm SO(3)}$ dynamical subgroup  chain,
provides a much more rapidly convergent basis for the expansion of the
eigenfunctions of an arbitrary collective model than does a basis that reduces the
chain ${\rm U(5)}\supset{\rm SO(5)}\supset{\rm SO(3)}$.  
For a collective model Hamiltonian that is SO(5)-invariant,  the
eigenfunctions and spectrum are readily computed using just the Lie algebra of
SU(1,1) as a spectrum generating algebra. Convergence is then  optimized by
selecting a value of $\beta_0$ at the minimum of the $\beta$
potential of interest.

For a general collective Hamiltonian, in which both the SU(1,1)
and SO(5) degrees are active,  expressions are needed for the SO(5) spherical
harmonics.
They can be constructed by the methods of Chacon {\em et al.} \cite{CMS}.
However,  I follow a simpler construction based on the observation that the SO(5)
spherical harmonics are an orthonormal basis for the Hilbert space
$\mathcal{L}^2(S_4)$ of square integrable functions on the four-sphere $S_4$ with
respect to the usual volume element \cite{BM}  given, in standard 
$(\gamma,\theta,\varphi)$ coordinates, by
\be {\rm d}v = \sin 3\gamma \, {\rm d}\gamma \, {\rm d}\Omega ,\label{eq:S4vol}\ee
where d$\Omega = \sin\theta\, {\rm d}\theta\, {\rm d}\varphi$.
A non-orthonormal basis for $\mathcal{L}^2(S_4)$ is simply constructed as shown in
sect.~\ref{sect:basiswfns}. 
This basis is then Gramm-Schmidt orthogonalized to give the SO(5) spherical
harmonics.
This  method will be deetailed and used to compute a subset of SO(5) Clebsch-Gordan
coefficients in a following paper \cite{RTR}.
Thus, the facility to diagonalize any Hamiltonian on the collective model Hilbert
space (\ref{eq:BMHilb}) and to compute, for example, $B$(E2) transition rates
between its eigenstates,  is limited only by the computational power available.
An illustrative calculation of the spectrum and E2 transitions for an axially
symmetric rotor-vibrator model is given in sect.~\ref{sect:results}. 

\section{Results for the basic SO(5)-invariant model}

\begin{figure}[ht]
\epsfig{file=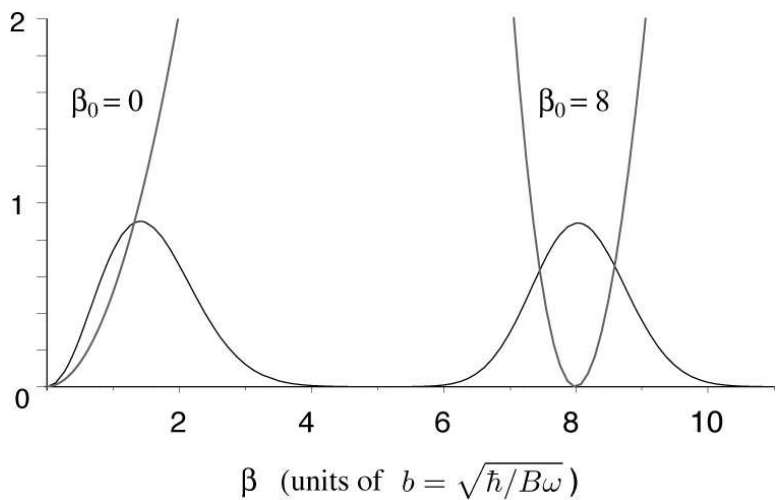, width=4.0 in}  
   \caption{The potential energy components of the Hamiltonian
(\protect\ref{eq:betaHam}) for $\beta_0=0$ and 8 together with the corresponding beta
wave functions multiplied by $\beta^2$. The potential is given in
units of $\hbar\omega$.
\label{fig:v0wfns}}  
\end{figure}

Figure \ref{fig:v0wfns} shows the potential energy function
\be V_{\beta_0}(\beta) = \frac{\hbar\omega}{2b^2} \left( \beta^2 +
\frac{\beta_0^4}{\beta^2}\right) 
\ee
 of the Hamiltonian
(\ref{eq:betaHam}) for $\beta_0 = 0$ and 8 together with the corresponding
ground-state $\beta$ wave functions.
The ratio of the $\beta$-width to the mean value of $\beta$, given  for these
wave functions, by $b/\beta_0$ is proportional to $1/\beta_0\sqrt{\omega}$. Thus, as
$\beta_0\sqrt{\omega}$ becomes large, the wave functions approach the Wilets-Jean
rigid-beta limit.  

\begin{figure}[ht]
\epsfxsize=3in
\centerline{\epsfbox{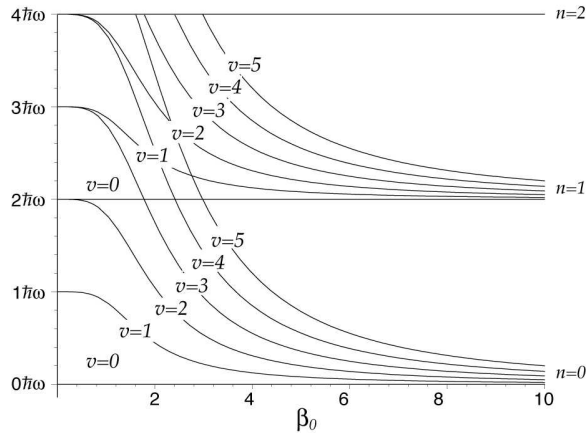}}
   \caption{Energies of the five-dimensional Davidson
model as a function of $\beta_0$. \label{fig:EDavidson}}  
\end{figure}

\begin{figure}[ht]
   \epsfig{file=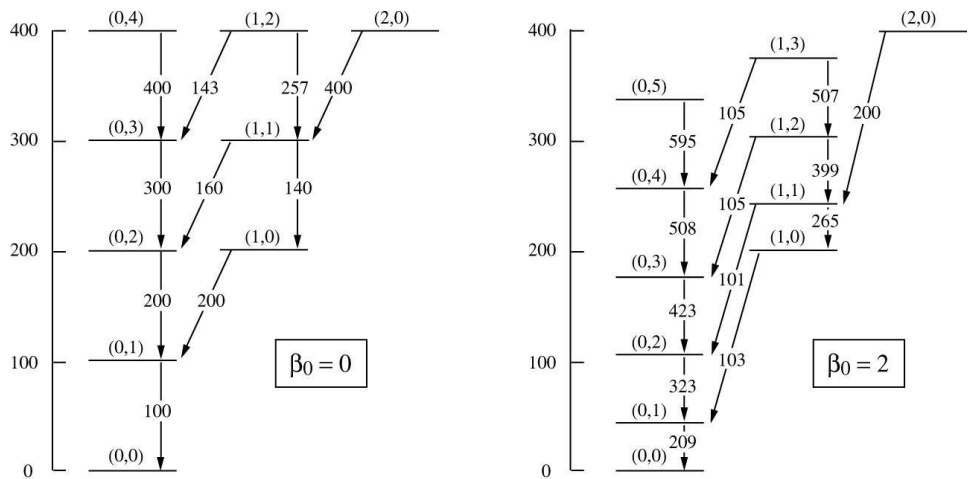, width=5.0 in}  
   \caption{Energy-levels and $B$(E2) transition rates for the Hamiltonian
(\protect\ref{eq:betaHam}) with $\beta_0=0$ and 2.  Energy levels are labeled by
their SU(1,1)$\times$SO(5) quantum numbers $(n,v)$.  Energies and $B$(E2) transition
rates are given in units such that the one-phonon $L=2$ excited (0,1) state  
has an excitation energy of 100 units and its $B$(E2) transition rate to the ground
state is also 100 units in the U(5) $(\beta_0=0)$  limit.
\label{fig:betaengies1}}   
\end{figure}

\begin{figure}[ht]
   \epsfig{file=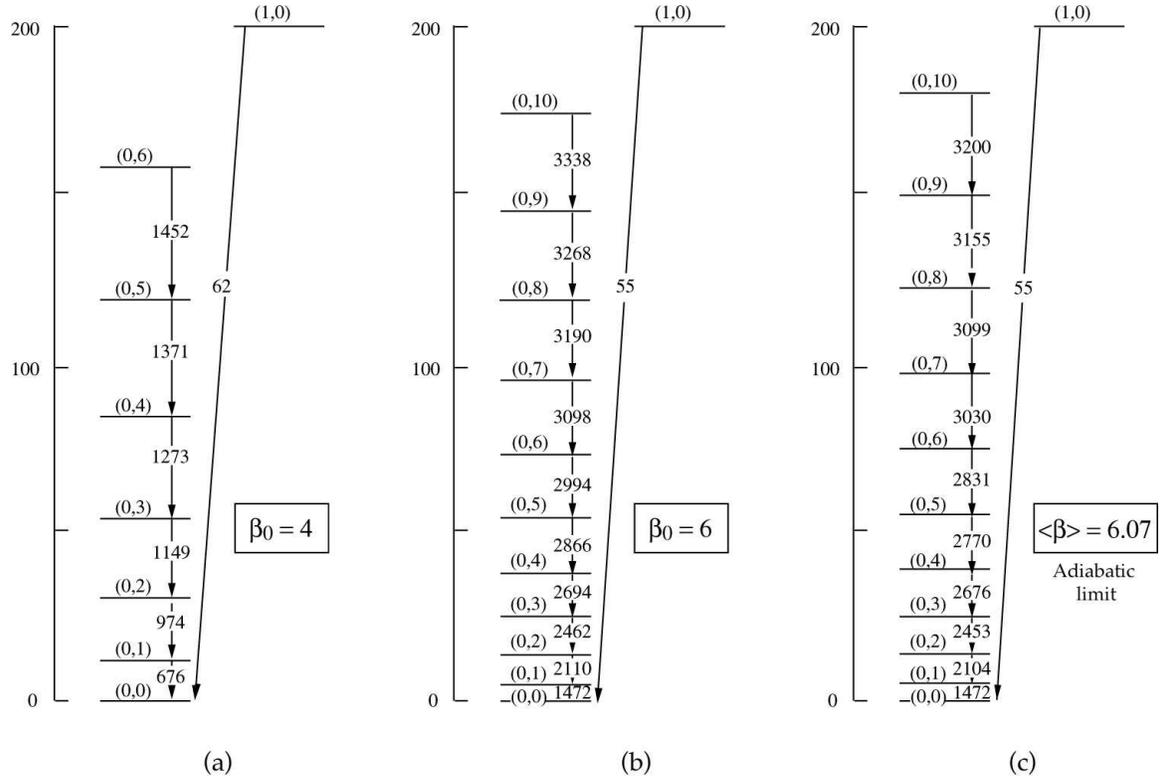, width=6.0 in}  
   \caption{Energy-levels and $B$(E2) transition rates for the Hamiltonian
(\protect\ref{eq:betaHam}) with $\beta_0=4$ and 6 in units  defined in the caption
to fig.\ \protect\ref{fig:betaengies1}.  For comparison, (c) is the
spectrum obtained by a simple adiabatic approximation with $<\hat\beta>=6.07$ as
described in the text.
\label{fig:betaengies2}}  
\end{figure}

The energy-level spectrum, given by eqns.~(\ref{eq:engylevels}) and
(\ref{eq:lambda}) is shown as a function of $\beta$ in fig.~\ref{fig:EDavidson} for
a range of values of the deformation parameter $\beta_0$. 
Fig.~\ref{fig:EDavidson}  shows that as $\beta_0\sqrt{\omega}\to\infty$ (the
Wilets-Jean limit)  the energy levels asymptotically approach a harmonic oscillator
sequence with the states of each asymptotic level comprising an infinite number of
states. Low energy levels for specific values of $\beta_0$ are shown in more detail
together with $B$(E2) transition rates in figs.\ \ref{fig:betaengies1} and
\ref{fig:betaengies2}.

$B$(E2) transition rates are computed as follows.
Let $\{ |nv\sigma\rangle \}$ denote the basis states corresponding to the wave
functions of eqn.\ (\ref{eq:basis}) with $\lambda_v$ related to $v$ by
eqn.~(\ref{eq:lambda}). The $B$(E2) transition rate between two SO(5) levels is then
defined in the standard way for the quadrupole operator $\hat Q$ by summing the
squared matrix elements of $\hat Q$ over the states of the final level and averaging
over  initial states.
Thus 
\bea B({\rm E}2; n_iv_i \to n_fv_f) &=& \sum_{\sigma_i\sigma_f M}
\frac{1}{d(v_i)} |\langle n_fv_f\sigma_f |\hat Q_M|n_iv_i\sigma_i\rangle |^2
\nonumber\\ &=& \frac{d(v_f)}{d(v_i)} |\langle n_fv_f |||\hat Q ||| n_iv_i\rangle|^2
\nonumber\\ &=& |\langle n_iv_i |||\hat Q ||| n_fv_f\rangle|^2 ,
\eea
where $d(v)$ is the dimension of the SO(5) irrep of seniority $v$ and   
$\langle n_fv_f |||\hat Q||| n_iv_i\rangle$ is an SO(5)-reduced matrix element of the
$v=1$ quadrupole tensor.
Since the wave functions for the states  $\{ |nv\sigma\rangle \}$ are products of
$\beta$ wave functions and spherical harmonics, the matrix elements of the  $\hat
Q_M = Z\hat \beta\hat q_M$ operators of eqn.\ (\ref{eq:Qop}) factor and are given by
\be \langle n_iv_i |||\hat Q ||| n_fv_f\rangle = Z \langle n_iv_i |\hat \beta|
n_fv_f\rangle
\langle v_i |||\hat q ||| v_f\rangle ,\ee
with obvious notation.
The $\hat \beta$ matrix element is readily evaluated for any value of $\beta_0$
while the
$\hat q$ matrix element is independent of $\beta_0$ and can be evaluated in the U(5)
($\beta_0=0$) limit.  

In the U(5) limit, the $\hat Q_M$ operator can be expressed in terms of $L=2$
harmonic-oscillator raising and lowering operators
\be \hat Q_M = Z  \frac{d^\dag_M + d_M}{\sqrt{2}} .\ee
It follows that  $\langle 0v|||\hat Q|||0v\rangle = 0$ and
\be \langle 0,v+1|||\hat Q|||0v\rangle = Z\sqrt{(v+1)/2}\ .\ee
Therefore, since the $\beta$ integral gives $\langle
0,v+1|\hat\beta|0v\rangle=\sqrt{(2v+5)/2}$, we infer that
\bea
&\langle v+1|||\hat q|||v\rangle = \displaystyle\sqrt{\frac{v+1}{2v+5}} , &\\
&\langle v|||\hat q|||v+1\rangle = (-1)^\phi\displaystyle\sqrt{\frac{v+1}{2v+5}\cdot
\frac{d(v+1)}{d(v)}} ,& \label{eq:qSO5}
\eea
where  the phase factor $(-1)^\phi$ depends on a choice of phase convention.
Thus,  for arbitrary $\beta_0$,
\bea  &B({\rm E}2; n'v\to n,v-1) = \displaystyle\frac{Z^2v}{2v+3}\langle
n'v|\hat\beta|n,v-1\rangle^2  ,&\\ 
&B({\rm E}2; n'v\to n,v+1) = \displaystyle\frac{Z^2(v+1)d(v+1)}{(2v+5) d(v)} 
\langle n'v|\hat\beta|n,v+1\rangle^2 .&\eea 
It is also follows from the identity  $\langle 0v|||\hat Q|||0v\rangle = 0$ that all
quadrupole moments are zero in this model as expected for a gamma-soft model.
For the present calculations, $Z^2$ was given the value $Z^2=200$ so that $B({\rm
E}2; 2_1\to 0_1) = 100$ in the U(5) ($\chi = 0$) limit.

It is seen from the figures that, as the value of $\beta_0$ is increased, excited
beta-vibrational bands separate and increase in energy.  In the Wilets-Jean limit,
in which
$\beta_0\sqrt{\omega}\to\infty$, the ratio of the beta-vibrational energy to the
lowest rotational excitation energy diverges.
As this limit is approached, a simple adiabatic approximation gives increasingly
good approximate solution to the results of the Hamiltonian (\ref{eq:betaHam}) as
illustrated by comparison of figs.\ 4(b) and 4(c).

The adiabatic approximation follows by expanding the expression for the energy
(\ref{eq:engylevels}) with $\lambda_v$ given by eqn.\ (\ref{eq:lambda}), for large
values of $\beta_0$,
\be E_{nv} = \left[ 2n + \frac{1}{2\beta_0^2} v(v+3) \right] \hbar\omega  + 
0\left(\frac{1}{\beta_0^4}\right) + {\rm const.},
\ee
and neglecting terms of order $1/\beta_0^4$ and higher.
This approximation corresponds to neglecting the centrifugal coupling between the
rotational and  beta-vibrational degrees of freedom.
Perturbation theory  shows that this approximation is good so long as the
rotational energy $v(v+3)/2\beta_0^2$, for the gamma-soft rotor, is small in
comparison with the beta-vibrational energy $\hbar\omega$.

\begin{figure}[ht]
   \epsfig{file=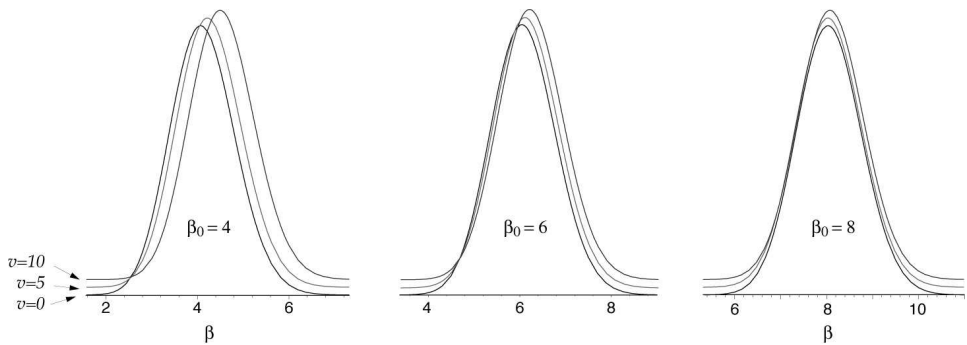, width=5.0 in}  
   \caption{Comparison of the beta components of the model wave functions for
$\beta_0=4,6,8$ and for $v=0$, 5, and 10.  It is seen that the wave functions are
shifted  to higher values of $\beta$ with increasing $v$ due to centrifugal
stretching for small values of $\beta_0$ but that as $\beta_0\sqrt{\omega}\to\infty$,
the adiabatic limit is approached in which this shifting becomes more and more
negligible for finite values of $v$.
\label{fig:betawfns}}  
\end{figure}

The effective decoupling of the beta-vibrational and rotational degrees of freedom
in the adiabatic limit, is evidenced by the independence of the beta wave
function on the seniority quantum number $v$ in this limit.  A comparison is made
between the $v=0$, 5, and 10 beta wave functions in fig.\ \ref{fig:betawfns} for
$\beta_0=4$, 6, and 8.  It is seen that centrifugal stretching is substantial for
$\beta_0=4$ but  negligible for $\beta_0 \approx 8$ and $v\lesssim 10$

\section{More general collective model Hamiltonians}

An extension of the above model to admit gamma
as well as deformed-beta equilibrium shapes, is given, for example, by adding a
gamma-dependent potential to the Hamiltonian (\ref{eq:betaHam});
\be \hat H = {\hbar^2\over 2B}\Big(- \nabla^2 +{\beta_0^4\over
\beta^2}\Big)+ \half B\omega^2\beta^2 + \hat V
\,. \label{eq:CMHam}
\ee
To illustrate, I consider the potential
\be V(\gamma) = -\chi \cos 3\gamma
\ee
which has an axially symmetric minimum.
This potential produces a very tractable model because, to within
a constant, $\cos 3\gamma$ is the SO(5) $v=3,L=0$ spherical harmonic.
An even simpler model is obtained by assuming a value of
$\beta_0\sqrt{\omega}$ for which  the $\beta$ and SO(5) degrees of freedom are
adiabatically decoupled.
The latter choice is not necessary but simplifies the calculations considerably.

\subsection{The adiabatic map}

For each value of the $\beta$ quantum number  $n$, the SO(5)
quantum number $v$ runs over the complete set of nonnegative integers
$v=0$, 1, 2, $\dots$.
Moreover, just as the set of SO(3) spherical harmonics $\{Y_{LM}\}$ spans the
Hilbert space $\mathcal{L}^2(S_2)$ of square integrable functions on the two-sphere,
so  the set of SO(5) spherical harmonics $\{\mathcal{Y}_{v\sigma M}\}$ spans the
Hilbert space
$\mathcal{L}^2(S_4)$    of square-integrable functions  on the four-sphere
(isomorphic to the factor space SO(4)$\backslash$SO(5)). It follows that
\be \mathcal{L}^2(S_4) = \bigoplus_{v=0}^\infty\, \Hb^{{\rm SO(5)}}_v \,.
\ee
Observe also that the SU(1,1) Hilbert spaces $\Hb^{{\rm SU(1,1)}}_v$ featuring in
eqn.\ (\ref{eq:BMHilb}) are all isomorphic to the common Hilbert space 
$\Hb^{{\rm SU(1,1)}}=\mathcal{L}^2(R^+)$ of functions of 
$\beta\in R^+$ (the positive radial line) that are square integrable with respect to
the measure $\beta^4\, {\rm d}\beta$.
Thus, we can define an {\em adiabatic map} from $\Hb^{{\rm SU(1,1)}}_v$ to the
isomorphic Hilbert space $\Hb^{{\rm SU(1,1)}}_0$ in which the
basis wave functions map
\be \phi_{nv} \mapsto \phi_{n0} \equiv \phi_{n} \,.\ee
Under this map
\be \Hb^{{\rm CM}} \to  \Hb^{{\rm SU(1,1)}}_0\otimes\mathcal{L}^2(S_4) \,.
\ee
The corresponding map of the Hamiltonian is given by the observation that, for
$\beta_0^4$ large compared to $(v+3/2)^2$,
\be \lambda_v = 1+\sqrt{(v + \textstyle {3\over 2})^2 + \beta_0^4} \quad\to\quad
\lambda_0  + {1\over 2\beta_0^2} v(v+3)\,.
\ee
Since $v(v+3)$ is an eigenvalue of the SO(5) Casimir operator   
$\hat\mathcal{C}^{\rm SO(5)}$, it follows that the adiabatic map sends the
Hamiltonian
$\hat H(\beta_0)$ to
\be \hat H(\beta_0)\to \hat H_0(\beta_0) +  {\hbar\omega\over
2\beta_0^2}\,\hat\mathcal{C}^{\rm SO(5)} \,,
\ee
where $\hat H_0(\beta_0)$ is the SU(1,1) Hamiltonian $\hat H(\beta_0)$ restricted to
the scalar $v=0$ representation on $\Hb^{{\rm SU(1,1)}}_0$.
Thus, the Hamiltonian $\hat H_0(\beta_0)$ gives a harmonic sequence of
$\beta$-vibrational states with energies  $\{n\hbar\omega, n=0,\dots, \infty\}$. 
The Hamiltonian $\hat H$ of eqn.\ (\ref{eq:CMHam}) likewise maps, in the adiabatic
limit, to
\be \hat H(\beta_0)\to \hat H_{\rm ad}(\beta_0) =\hat H_0(\beta_0) + 
{\hbar\omega\over 2\beta_0^2}\,\hat\mathcal{C}^{\rm SO(5)}+\hat V. \label{eq:}\ee
I consider the Hamiltonian
\be \hat \mathcal{H}(\chi)= \hat\mathcal{C}^{\rm SO(5)} -\chi \cos 3\gamma
\label{eq:Hchi} \ee 
as a component of the adiabatic Hamiltonian
\be H_{\rm ad}(\beta_0,\chi) =\hat H_0(\beta_0) +  {\hbar\omega\over
2\beta_0^2}\,\left[\hat\mathcal{C}^{\rm SO(5)}-\chi \cos 3\gamma\right]
.  \label{eq:Hbetagamma}
\ee
Then, if the energy levels of $\hat\mathcal{H}(\chi)$ are given by $\{
\mathcal{E}_{\alpha L}\}$, the energy levels of $H_{\rm ad}(\beta_0,\chi)$ are
given by
\be E_{n\alpha L} = n\hbar\omega + {\hbar\omega\over
2\beta_0^2}\,\mathcal{E}_{\alpha L}.
\ee

\subsection{Basis wave functions and matrix elements} \label{sect:basiswfns}

The following construction makes substantial use of  the methods of
Chacon, Moshinsky, and Sharp  \cite{CMS} but is conceptually and
computationally simpler.

The primary observation is that the Hilbert space of square integrable functions on
the four-sphere is spanned by  polynomials of the elementary $v=1, L=2$ quadrupole
moment functions:
\be q_M(\gamma,\Omega) = \cos\gamma \,\mathcal{D}^2_{0,M}(\Omega) +
\frac{1}{\sqrt{2}} \sin\gamma \left[
\mathcal{D}^2_{2,M}(\Omega)+\mathcal{D}^2_{-2,M}(\Omega)\right] . \label{eq:q}
\ee
To construct a basis, I start by forming a minimal set of angular-momentum-coupled
wave functions, with angular momentum projection $M=L$, from which a
complete set of $M=L$ wave functions can be generated
by taking products of the wave functions in this minimal set. A huge
advantage is gained by proceeding in this way because all the
wave functions generated  have good angular momentum and form a complete
set of highest-weight wave functions for the SO(3) irreps in $\mathcal{L}^2(S_4)$.

A suitable minimal set of wave functions is  found by examination of the
angular momentum content of the  $\mathcal{L}^2(S_4)=\bigoplus_v\, \Hb^{({\rm
SO(5)})}_v$ space. Knowing the SO(5)$\to$SO(3) branching rules for the irreps
appearing in the collective model \cite{Brules}, it is possible to arrange the
angular-momentum states of $\mathcal{L}^2(S_4)$ into $K$ bands, each
having the same sequence of angular-momentum states as those of  axially-symmetric
rotor bands of the same
$K$, and with the sequence of
$K$ bands being in one-to-one correspondence with those of a sequence of
gamma-vibrational bands. More precisely, the band-heads of the $K$ bands appear with
increasing seniority
$v$ in the sequence
\be \begin{array}{cccccccccccccc}
v= &0&1&2&3&3&5&6&7&8&9&10& \cdots \\
K= & 0&&2&&4&&6&&8&&10&\cdots \\
   &&&&0&&2&&4&&6&& \cdots \\
&&&&&&&0&&2&&4& \cdots \\
\end{array} 
\ee
This band structure, albeit with different energies, is precisely that of the
axially-symmetric rotor-gamma-vibrator model (i.e., no beta-vibrational bands) as can
be seen in the spectrum of the Hamiltonian $\hat\mathcal{H}(\chi)$ for $\chi=0$
of fig.\ \ref{fig:chi0}.
It is  now seen that a complete set of $M=L$ coupled polynomials in ${q_M}$
 is generated by taking multiple products of the four generating functions
\bea &\Phi_{0022} \propto q_2, \\
&\Phi_{0222} \propto [q\otimes q]_{22},\\ 
&\Phi_{1000} \propto [q\otimes q\otimes q]_0 ,\\
&\Phi_{0233} \propto [q\otimes q\otimes q]_{33}.
\eea
Thus,  these  functions generate a linearly-independent basis of $M=L$ polynomials
\be
\Phi_{tKLL} = [\Phi_{0022}]^{n_1} [\Phi_{0222}]^{n_2} [\Phi_{1000}]^{t}
[\Phi_{0233}]^{n_3} , \label{eq:polys}
\ee
with
\be K = 2n_2+ 2n_3, \quad L= 2n_2+2n_2+3n_3 ,\quad n_3=0 \; {\rm or} 1. \ee
If a state belonging to a $K$ band is labeled by the index
$t=i-1$ when it appears in the $i$'th occurence of $K$ (e.g., $t=0$ for states of the
lowest $K=0$ band and $t=1$ for states of the first excited $K=0$ band), then the
polynomials of eqn.\ (\ref{eq:polys}) are in one-to-one correspondence with the
states labeled by the $K$-band system  given above; 
$t$ is then the  number of $\Phi_{1000}$ zero-coupled triplets as in the  basis
construction of \cite{CMS}.

The polynomials $\{\Phi_{tKLM}\}$ do not form an orthonormal basis.
However, their overlaps are readily evaluated using the inner product for
$\mathcal{L}^2(S_4)$ defined by the standard volume element 
(\ref{eq:S4vol}). 
Moreover, since a polynomial $\Phi_{tKLL}$ of degree $N$ in $q$
does not contain admixtures of wave functions of seniority greater than $N$,  it is a
simple matter to sequentially Gramm-Schmidt orthogonalize these polynomial
functions  to obtain an orthonormal basis of wave functions 
$\{\Psi_{vKLM}\}$ of good SO(5) seniority $v$.
(Note that the label $K$ of the orthonormal basis is not a good quantum number;
like the Vergados $K$ label of the SU(3) model it is just a convenient label which
makes a useful correspondence with the rotor model.) The procedure
will be described in detail in ref.~\cite{RTR}. 

In manipulating $\mathcal{L}^2(S_4)$ wave functions, it is convenient to expand
them in the form
\be \Psi (\gamma,\Omega) = \sum_{\kappa\geq 0}^{\rm even} F_\kappa(\gamma)
\sqrt{2L+1\over 16\pi^2 (1+\delta_{\kappa0})} 
\left[ \mathcal{D}^L_{\kappa M}(\Omega) + (-1)^L\mathcal{D}^L_{-\kappa M}(\Omega)
\right] .
\ee
Thus, a wave function $\Psi\in\mathcal{L}^2(S_4)$ is defined by a set of functions
$\{ F_0(\gamma), F_2(\gamma), \dots \}$.
In particular, the above generating functions are given by 
\bea &\Phi_{0022} \sim \{ \cos\gamma,\sin\gamma\}, \\
&\Phi_{0222} \sim  \{ \cos 2\gamma,\sin 2\gamma\},\\ 
&\Phi_{1000} \sim  \{ \cos 3\gamma\},\\
&\Phi_{0233} \sim  \{ 0,\sin 3\gamma\}.
\eea  

The overlaps of wave functions are given in this representation by
\be \langle \Psi_{LM}|\Psi'_{L'M'}\rangle = \delta_{LL'} \delta_{MM'}
 \sum_{\kappa\geq 0}^{\rm even} \int F^*_{L\kappa}(\gamma)F'_{L\kappa}(\gamma)\, \sin
3\gamma\, {\rm d}\gamma
\ee
and matrix elements of the interaction $\cos 3\gamma$ by
\be \langle \Psi_{LM}|\cos 3\gamma|\Psi'_{L'M'}\rangle = \delta_{LL'}
\delta_{MM'}
 \sum_{\kappa\geq 0}^{\rm even} \int F^*_{L\kappa}(\gamma)F'_{L\kappa}(\gamma)\, \cos
3\gamma\sin 3\gamma\, {\rm d}\gamma
\ee
In the present calculation, the needed matrix elements and overlap integrals were
evaluated analytically using the interactive mathematics computer program `Maple'. 

\section{E2 transition rates and quadrupole moments}

When the adiabatic approximation is valid, the  matrix elements of the quadrupole
operators $\hat Q_M = Z\hat \beta\hat  q_M$ factor
\be \langle \Psi_L\|\hat Q\|\Psi_{L'}\rangle = Z \langle\beta\rangle \langle \Psi_L\|
\hat q\|\Psi_{L'}\rangle .\ee
Moreover the expectation value $\langle \beta\rangle$, which varies
continously with $\beta_0$ can itself be treated as a free parameter. Thus, it
only remains to compute the reduced matrix elements of
$\hat q$. This is straightforward using the wave functions contructed  according to
the methods of the previous section.
For example, in the SO(5) ($\chi=0$) limit, the $v=0,L=0$ and $v=1,L=2$ states of the
orthornormal set
$\{ |vKLM\rangle\}$ have wave functions
\bea &|0000\rangle \sim \displaystyle\sqrt{\frac{3}{16\pi^2}}  \\
&|102M\rangle \sim \displaystyle\sqrt{\frac{15}{16\pi^2}} \left[ \cos\gamma\,
\mathcal{D}^2_{0M}(\Omega) + \frac{1}{\sqrt{2}}\sin\gamma\,
\left(\mathcal{D}^2_{2M}(\Omega)+\mathcal{D}^2_{-2,M}(\Omega)\right) \right].
\eea
Thus, with $\hat q_M$ given by eqn.\ (\ref{eq:q}), 
\be \langle 102\|\hat q\| 000\rangle = 1 \ee
and  
\be B({\rm E}2: L_i\to L_f) = Z^2 \langle\beta\rangle^2 {|\langle L_f\|
\hat q\|L_i\rangle|^2 \over 2L_i+1} , \ee
it follows that
\be B({\rm E}2: 102\to 000) = \frac{1}{5} Z^2 \langle\beta\rangle^2 .\ee

The calculated results can be used with arbitrary values of the factor
$Z\langle\beta\rangle$. The transition rates shown in the figures are in units such
that, in the SO(5) limit  (in which $\chi=0$ and $v$ is good quantum number; cf.\
fig.\ \ref{fig:chi0}) 
\be B({\rm E}2: 102\to 000) = 100.\ee
$B$(E2) values  in these units are given explicitly by the expressions
\be B({\rm E}2: L_i\to L_f) \equiv 100\, \frac{B({\rm E}2: L_i\to L_f)}{B({\rm E}2:
102\to 000)} = 500\,{|\langle L_i\|\hat q\|L_f\rangle|^2
\over 2L_i+1}.
\ee
Quadrupole moments are given in the corresponding units by
\be Q(L) = \sqrt{500}\, (LL,20|LL) \frac{\langle L\|\hat q\|L\rangle}{\sqrt{2L+1}} ,
\ee
where $(LL,20|LL)$ is a standard SO(3) Clebsch-Gordan coefficient.

In the SO(5) limit, when $\chi=0$ and $v$ is a good quantum number, 
\be \frac{\langle v+1,L_i\|\hat q\|v,L_f\rangle}{\sqrt{2L_i+1}} = 
(vL_f,12\|v+1,L_i)\, \langle v+1 |||\hat q|||v\rangle 
\ee
where $(vL_f,12\|v+1,L_i)$ is a reduced SO(5) Clebsch-Gordan coefficient and
$\langle v+1 |||\hat q|||v\rangle$ is the SO(5)-reduced matrix element given by
eqn.\ (\ref{eq:qSO5}). Thus, in the SO(5) limit,
\be \frac{B({\rm E}2: v+1,L_i\to vL_f)}{ B({\rm E}2: 102\to 000)}
=\frac{5(v+1)}{(2v+5)} (vL_f,12\|v+1,L_i)^2.
\ee
Low-lying values of these ratios are given in Table \ref{tab:E2}.
Quadrupole moments vanish in the SO(5) limit because $\hat q$ is a $v=1$ tensor
operator and, hence, its matrix elements satisfy the well-known $\Delta v = \pm 1$
selection rule.

\begin{table}
\caption{\label{tab:E2}Values of the 
$B({\rm E}2: vL_i\to v-1,L_f)/ B({\rm E}2: 1 2_1\to 00_1)$ ratio
computed exactly as rational fractions.}
\begin{ruledtabular}
\begin{tabular}{r|ccccc}
$v\quad L_i$ & $L_f=L-2$& $L-1$ & $L$ & $L+1$ & $L+2$ \\ \hline
$1\quad 2_1$ & 1 \\ \hline
$2\quad 4_1$ & 10/7 \\
$\qquad 2_2$ & && 10/7 \\  \hline
$3\quad 6_1$ & 5/3 \\
$\qquad 4_2$ & 55/63 && 50/63 \\
$\qquad 3_1$ & & 25/21 && 10/21 \\
$\qquad 0_2$ & &&&& 5/3 \\ \hline
$4\quad 8_1$ & 20/11 \\
$\qquad 6_2$ & 150/121 && 70/121 \\
$\qquad 5_1$ & 21/22 &105/242 & & 52/121 \\
$\qquad  4_3$ & &26/26 & 910/1089 & & 64/3267 \\
$\qquad 2_3$ & 2/3 & & & 5/6 & 7/22\\\hline
$5\quad 10_1$ & 25/13 \\
$\qquad 8_2$ & 19/13 && 6/13\\
$\qquad 7_1$ & 120/91 & 3/13 && 34/91\\
$\qquad 6_3$ & 16065/20449 &340/1001 & 1224/1573 && 256/13013\\
$\qquad 5_2$ & & 189/143 & 6/13 & 20/143 \\
$\qquad 4_4$ &75/77 && 1728/11011 &7/11 & 245/1573\\
$\qquad 2_4$ & && 10/7&&45/91\\ \hline
$6\quad 12_1$ & 2\\
$\qquad 10_2$& 92/57 &&22/57 \\
$\qquad 9_1$ & 55/36& 11/76 & 56/171\\
$\qquad 8_3$ & 1254/1105 & 19/117 & 847/1235 && 640/37791\\
$\qquad 7_2$ &68/91 & 22/35 & 121/273 & 19/105\\
$\qquad 6_4$ &0 & 32300/22841 &4598/8785 & 1292/22841 & 52/8785\\
$\qquad 6_5$ & 1004/845 &3136/212095 &544/3263 &91125/169676 &15827/169676 \\
$\qquad 4_5$ & 17/26 &&34/39 &17/78 & 10/39\\
$\qquad 3_2$ && 15/14 && 22/35 & 3/10\\
$\qquad 0_3$ &2\\
\end{tabular}
\end{ruledtabular}
\end{table}

\section{Results} \label{sect:results} 

\begin{figure}[htp]
   \epsfig{file=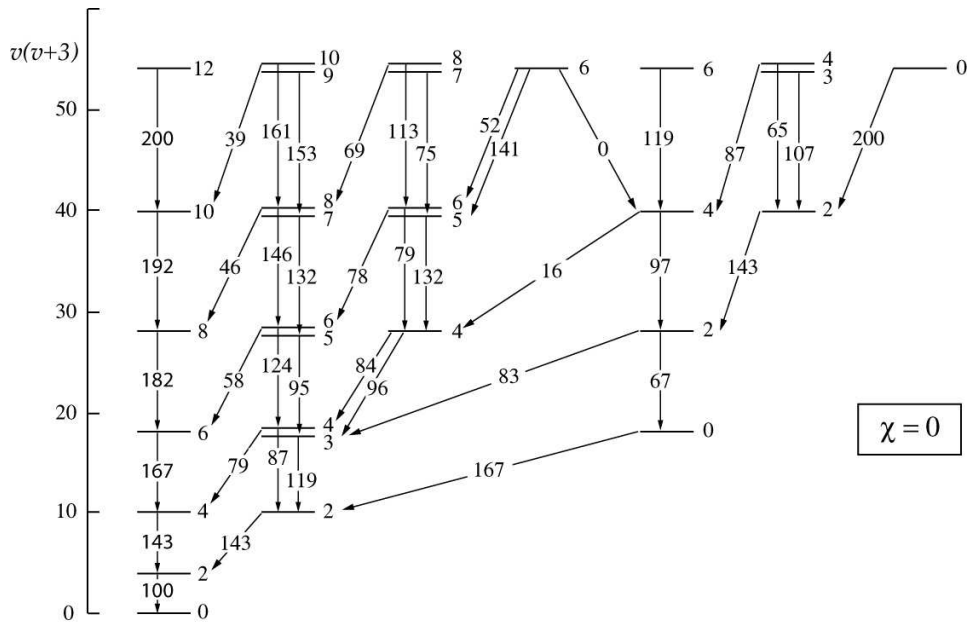, width=5 in}  
   \caption{The low-energy spectrum of the Hamiltonian (\ref{eq:Hchi}) in the SO(5)
limit in which seniority $v$ is a good quantum number. 
Some pairs of degenerate energy levels are separated slightly for clarity.
The $B$(E2) transition rates
shown are in units such that $B({\rm E}2; 12_1\to 00_1)= 100$. A more complete set of
transition rates is given (precisely) in table \ref{tab:E2}.  Quadrupole
moments vanish for states of good seniority.
\label{fig:chi0}}  
\end{figure}

\begin{figure}[htp]
   \epsfig{file=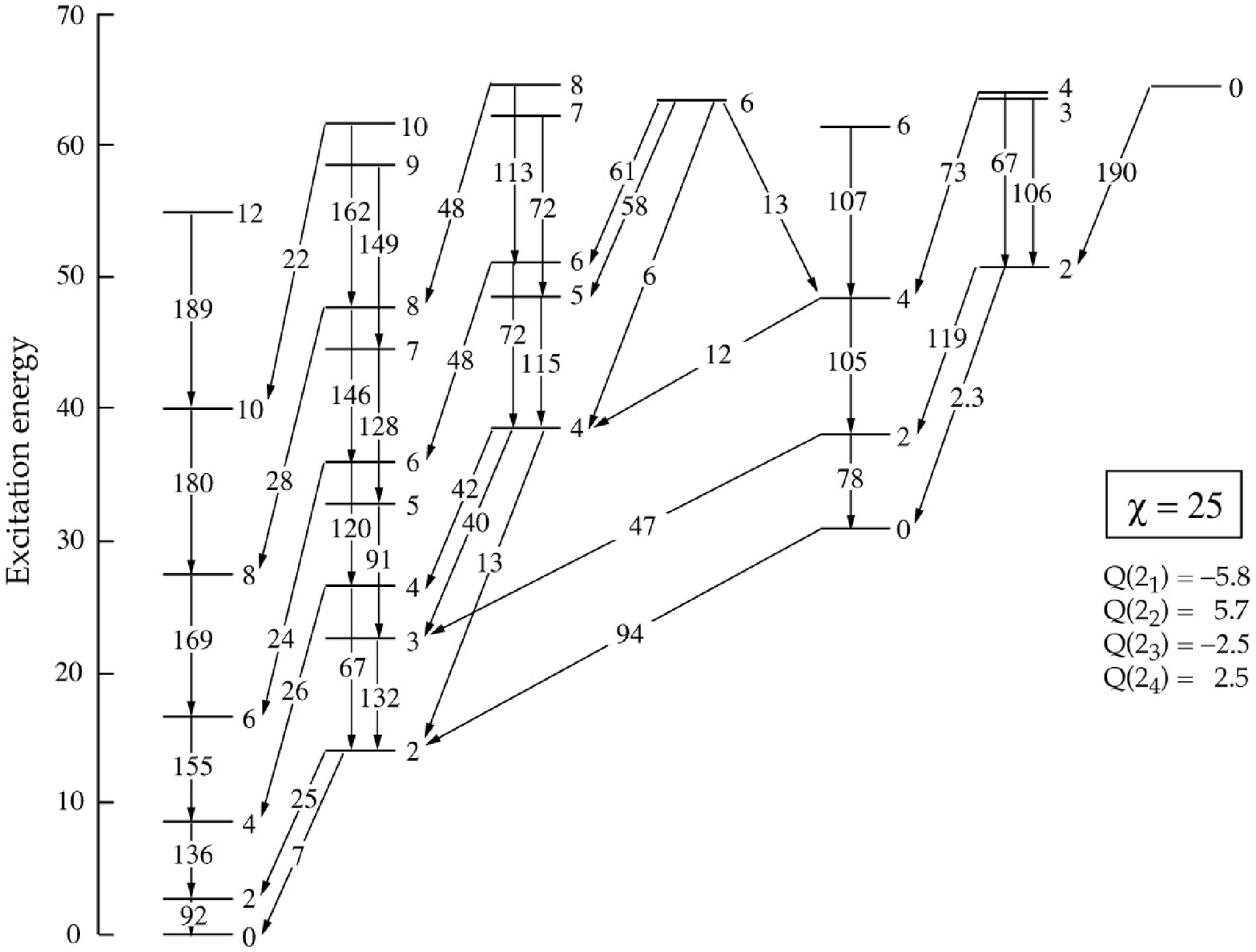, width=5 in}  
   \caption{The low-energy spectrum and $B$(E2) transition rates for the Hamiltonian
(\ref{eq:Hchi}) with $\chi = 25$.  The units are as defined in the caption to fig.\ 
\ref{fig:chi0}.  For $\chi$ non-zero, seniority is no longer a good quantum
number and quadrupole moments are no longer zero.
\label{fig:chi25}}  
\end{figure}

\begin{figure}[htp]
   \epsfig{file=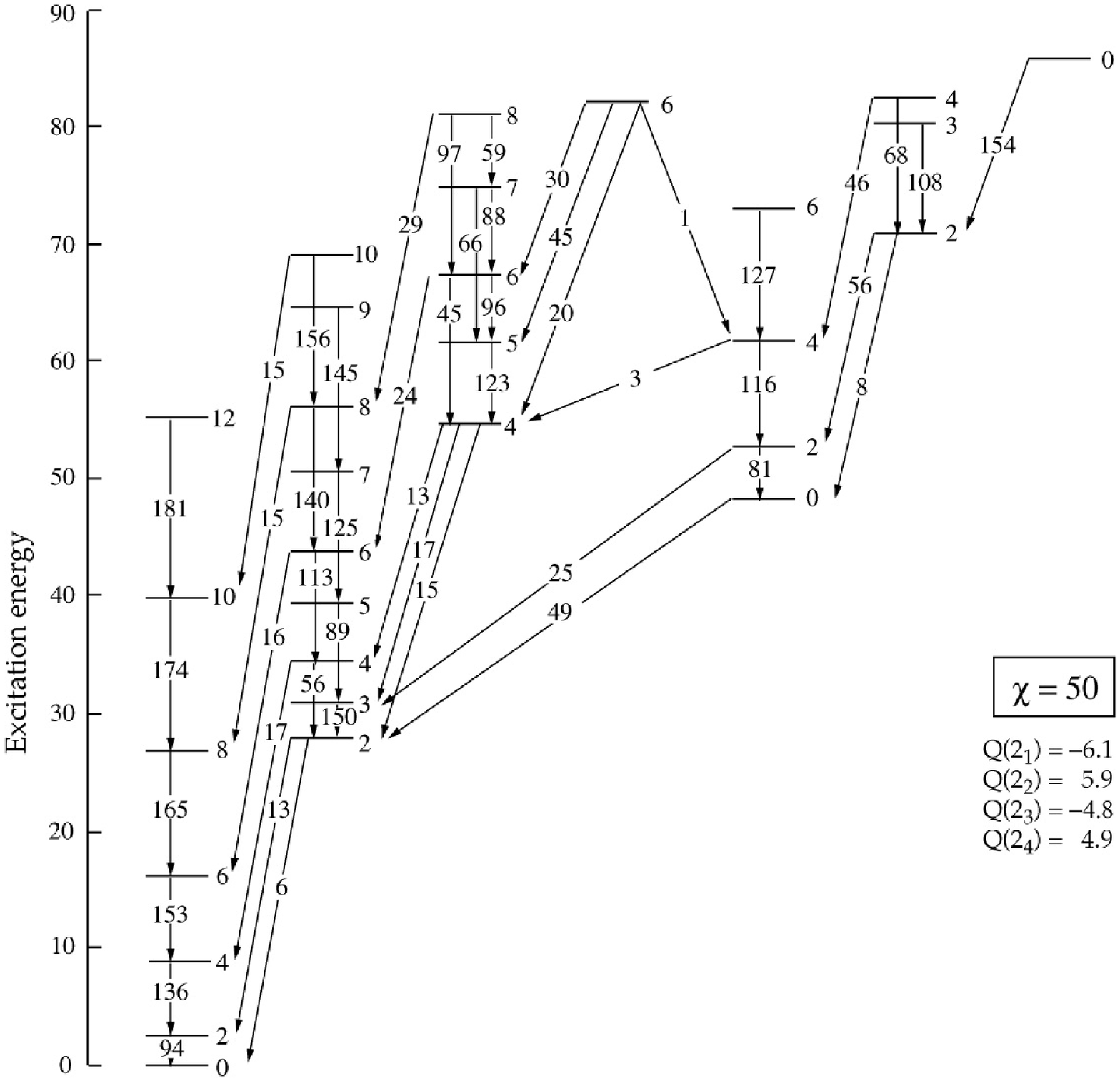, width=5 in}  
   \caption{The low-energy spectrum and $B$(E2) transition rates for the Hamiltonian
(\ref{eq:Hchi}) with $\chi = 50$.  The units are as defined in the caption to fig.\ 
\ref{fig:chi0}.  For $\chi$ non-zero, seniority is no longer a good quantum
number and quadrupole moments are no longer zero.
\label{fig:chi50}}  
\end{figure}

The low-energy spectrum of the Hamiltonian (\ref{eq:Hchi}) is shown for $\chi =0$,
25 and 50 in figs.\ \ref{fig:chi0}, \ref{fig:chi25}, and \ref{fig:chi50},
respectively.
The results were obtained by diagonalization of the Hamiltonian in a basis of 12
states for each angular momentum; it was ascertained that this number was sufficient
to obtain results of the desired accuracy for $\chi\leq 50$. It can be seen from
these figures that the spectrum and E2 transition rates progress with increasing
$\chi$ from those of the Wilets-Jean gamma-soft model (at $\chi=0$) to those of the
adiabatic axially-symmetric rotor-vibrator model. In particular, the $K$ bands
acquire  the characteristics of an axially-symmetric rotor model and sequences of
excited gamma-vibrational bands emerge.  This is most evident in the sequence of
$K=2$, 4 and 6 bands, which appear as one-, two-, and three-phonon gamma-vibrational
bands, and the move of the first excited $K=0$ band-head towards the energy of the 
two-phonon $K=4$ gamma-vibrational energy.

\begin{figure}[htp]
   \epsfig{file=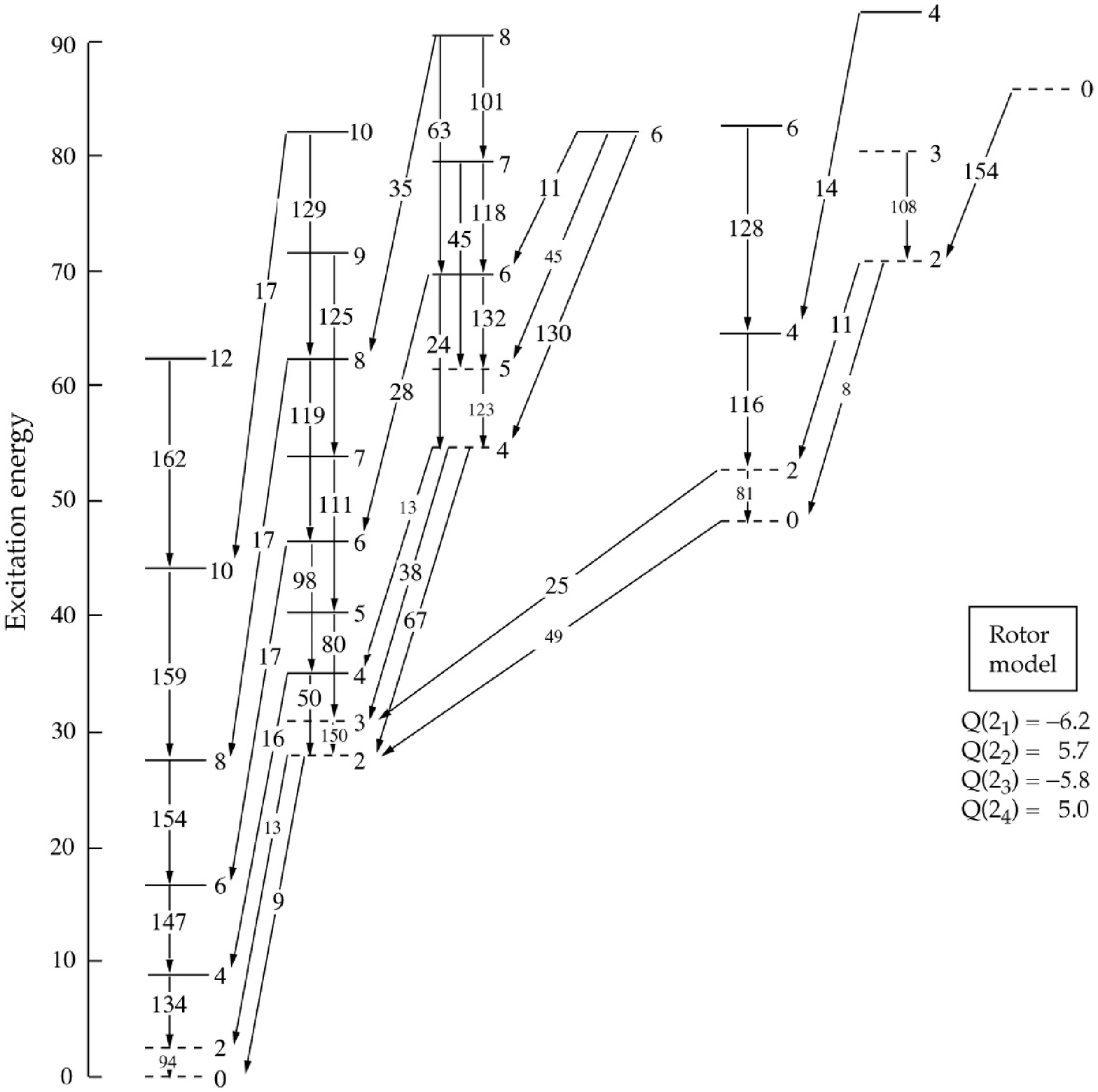, width=5 in} 
   \caption{The low-energy spectrum and $B$(E2) transition rates given by the
axially symmetric rotor  model with band-head energies and moments of inertia
adjusted so that the dashed energy levels are fitted to those of fig.\
\ref{fig:chi50}.  The intrinsic quadrupole moments of the model were also adjusted
to fit the E2 transition rates shown in smaller type.
\label{fig:RM}}  
\end{figure}

It is instructive to compare the results of fig.\ \ref{fig:chi50} with the sequence
of rotational bands expected in the standard axially symmetric rotor model shown in
fig.\ \ref{fig:RM}. 

The comparison shows that the rotational dynamics are adiabatic relative to the
gamma-vibrations for the lowest energy states but that significant departures become
evident at higher energies.
Increasing the value of $\chi$  would surely reduce the
rotational-vibrational coupling and give results closer to those of the adiabatic
rotor model.  However, by the same token, the gamma-vibrational
bands would rise to higher energies.
It is of interest to see if some of the effects of rotation-vibration coupling, e.g.\
the non-rotational behaviour of some of the  E2 transitions in the two-phonon $K=4$
gamma band, are realized in rotational nuclei having low-energy gamma bands.
It would also be interesting to see if the results of fig.\ \ref{fig:chi50} can be
reproduced more accurately in the adiabatic rotor model by taking the
rotation-vibration coupling into account in perturbation theory.

\section{Summary}

The Wilets-Jean model is an algebraic model with an 
${\rm [R^5]SO(5)} \supset {\rm SO(5)} \supset {\rm SO(3)}$
dynamical subgroup chain.
However, as observed above, the states that reduce this dynamical chain are
strictly zero-width beta-rigid states.
Complete rigidity is unphysical.
Nevertheless, it is possible for the WJ model to work  well
even when the beta fluctuations of physical states are relatively large.  
This is because, it is hard to measure the beta-width of physical states; indeed it
is invariably unclear as to the nature of excited $K=0$ excitations.
Many intrinsic excitations of a nucleus can give rise to excited $K=0$ bands and
even the collective model has two-phonon $K=0$ gamma bands.

To avoid the problems associated with the beta-rigidity of the WJ model, we
considered in this paper a more physical  collective model which, in its gamma-soft
limit, has an 
${\rm SU(1,1)\times SO(5)} \supset {\rm SO(5)} \supset {\rm SO(3)}$  dynamical
subgroup chain. For large values of the parameter
$\beta_0\sqrt{\omega}$, the low-energy states of this model acquire properties that
are essentially indistinguishable from those of the less physical WJ model. In the
WJ limit of this model, the beta vibrational frequency is infinite for a finite
value of $\beta_0$ and, hence, beta vibrational excitations are not observable. 
However, neither an infinite beta-vibrational frequency nor a zero
beta-width for collective wave function is necessary to obtain the results of the WJ
model. All that is needed is an adiabatic decoupling of the rotational and
beta-vibrational degrees of freedom.
The signature of such decoupling is that the beta wave functions in the ${\rm
SU(1,1)\times SO(5)}$ model become essentially independent of the excitation energy
for the range of energies of interest.  That such is the case when the
rotational energies are small compared to the one-phonon beta-vibrational energy is
illustrated in fig.\ \ref{fig:betawfns}.

Relaxing the rigidity of the WJ model and replacing its ${\rm [R^5]SO(5)}$ dynamical
group with ${\rm SU(1,1)\times SO(5)}$, results in a more physical
collective model.  More importantly, it makes it possible to construct a basis
in which arbitrary collective model Hamiltonians can be expanded in a rapidly
convergent manner.
For example, the solutions of the Hamiltonian (\ref{eq:Hbetagamma}) given in figs.\
\ref{fig:chi0}-\ref{fig:chi50} were obtained by diagonalization of
$12\times 12$ matrices.  For values of $\chi > 50$, the dimensions would have to be
increased to ensure accurate results.  But clearly, much larger dimensions can be
accommodated with available computers.

\appendix
\section{The five-dimensional Kratzer model}

The five-dimensional Kratzer model,  proposed by
Fortunato and Vitturi \cite{FV}, has a similar algebraic solution to that of the
Elliott-Evans-Park model \cite{EEP86}.
A straightforward extension of the  methods used for the hydrogen atom 
to the Hamiltonian
\be \hat H = -{\hbar^2\over 2B}\nabla^2 - {k\over \beta}  \,,
\ee
of a five-dimensional Coulomb problem gives the energy levels
\be E_{n\tau} = -{ k^2B\over 2\hbar^2\nu_{n\tau}^2} \,, \quad
\nu_{n\tau} =  n + \lambda_\tau/2\,,
\label{eq:EK}\ee 
and wave functions
\be \Psi_{n\tau}(\beta) = \sqrt{n!  \over
\Gamma(n+\lambda_\tau) 2\nu^6_\tau b^5}\ 
\Big( {\beta\over b\nu_\tau}\Big)^{(\lambda_\tau - 4)/2}
\exp \Big( - {\beta\over 2b\nu_\tau}\Big) \,
L^{(\lambda_\tau-1)}_n \Big(  {\beta\over b\nu_\tau}\Big)\,
\mathcal{Y}_{\tau\sigma}(\omega) , \label{eq:PsiK}
\ee
where now $\lambda_\tau = 2\tau+4$ and $b = \hbar^2/ (2kB)$.

These results are obtained 
by use of the SU(1,1) Lie algebra spanned by the operators
\be \hat Z_1 = -2\beta\nabla^2 , \quad \hat Z_2 = 2\beta , \quad \hat Z_3 =
-2i  (q\cdot \nabla + \textstyle{5\over 2})\,.
\ee
Again it is seen that, the commutation relations of this Lie
algebra are unchanged if $\hat Z_1$ is replaced by
\be \hat Z_1' = 2\beta\Big(-\nabla^2 + {\beta_0^4\over
\beta^2}\Big)
\,.
\ee
It follows that the above results for the Kepler Hamiltonian, extend
immediately to the five-dimensional Kratzer model with Hamiltonian
\be \hat H(\varepsilon) = {\hbar^2\over 2B}\Big(- \nabla^2 +{\beta_0^4\over
\beta^2}\Big) - {k\over \beta} \,.\ee
However, now  the relationship $\lambda_\tau = 2\tau + 4$ is replaced by
\be \lambda_\tau = 1+2\sqrt{(\tau + \textstyle {3\over 2})^2 + \beta_0^4}
\,.\ee
This follows because the  replacement $\hat Z_1 \to \hat Z_1'$ in the SU(1,1)
Lie algebra modifies the value of the SU(1,1) Casimir invariant from
the value
$\lambda_\tau (\lambda_\tau -2) = 4\tau (\tau+3) + 8$ to
the value
\be \lambda_\tau (\lambda_\tau -2) = 4\tau (\tau+3) + 8 + 
4\beta_0^4\, .\ee

 \end{document}